# Against a proposed alternative explanation of the Aharonov-Bohm effect

**Murray Peshkin**

Physics Division, Argonne National Laboratory, Argonne, IL 60439, USA

Email: peshkin@anl.gov

**Abstract.** It has been suggested from time to time that the Aharonov-Bohm effect is somehow a consequence of a classical electromagnetic field phenomenon involving energy that is temporarily stored in the overlap between the external field and the field of which the beam particle is the source. That idea was shown in the past not to work for some models of the source of the external field. Here a more general proof is presented for the magnetic AB effect to show that the overlap energy is always compensated by another contribution to the energy of the magnetic field in such a way that the sum of the two is independent of the external flux. Therefore no such mechanism can underlie the Aharonov-Bohm effect.

## 1. Introduction and conclusion

The Aharonov-Bohm (AB) effect is a feature of standard quantum mechanics that has no analog in classical physics. The motion of a charged particle may be influenced by externally fixed electromagnetic fields confined to regions from which the particle is rigorously excluded. That phenomenon is now understood to imply a previously unanticipated nonlocality in the action of the Maxwell fields on electric charges and it has provided a direct demonstration of the physical reality of the gauge fields.

When the AB effect was first announced [1] in 1959 it stimulated widespread surprise and much disbelief. Numerous attempts were published either to show that Aharonov and Bohm were simply wrong and the AB effect does not exist in the theory, or to eliminate that effect by changing quantum mechanics. Other proposed explanations attempted to ascribe the AB effect to classical mechanisms that did not require nonlocality in the electromagnetic interaction. With one exception, all of those erroneous ideas died out within a few years.

The lone exception is an idea based on classical physics. The moving charged particle is itself the source of a time-dependent electromagnetic field. That field may penetrate the domain of the external field, where the particle itself cannot go. The energy density in the total field is proportional to the squared sum of the two fields and the cross term in

Against a proposed alternative explanation of the Aharonov-Bohm effect that squared sum provides a time-dependent energy proportional to the external field. That energy must come from somewhere. The charged particle must be involved and the quantum mechanical consequences for its motion may somehow be affected in a way that depends upon the external field it cannot enter. If so, it is at least possible that no nonlocal interaction with the external electromagnetic field is required.

Simple implementations of that idea were shown not to work for the magnetic AB effect almost fifty years ago [1,2] but the idea apparently remains an attractive one because it has reappeared in various forms involving the magnetic or the electric AB effect from time to time until the present [3-5]. My purpose here is to show in a general way that, independently of any model of the source of the external field and independently of any details of the experiment, the basic concept underlying that idea is incorrect classical physics. No such time-varying energy of the total magnetic field within the domain of the external field can depend upon the external magnetic flux so no such mechanism can contribute to the Aharonov-Bohm effect. The same reasoning can be applied to models that make use of electric field energy in the electric AB effect but that is not done here because the case of the magnetic AB effect is simpler and exposes the main point more clearly, and because any one example suffices to show that the nonlocality revealed by the AB effect cannot be avoided by invoking the proposed classical mechanism.

Section 2 below contains a brief review of the magnetic AB effect. Section 3 describes the main idea behind the proposed alternative explanation and outlines previous more limited proofs that it is based on incorrect classical physics. In Section 4 a new, more general proof is given to show that no explanation based the idea of the proposed explanation of the AB effect can work. In Sections 2 and 3, it is assumed for simplicity that the geometry of the external magnetic field and its source is cylindrical. That assumption, which does not apply to some of the experiments, is removed at the end of Section 4.

Against a proposed alternative explanation of the Aharonov-Bohm effect

**2. The magnetic Aharonov-Bohm effect**
Figure 1 illustrates an idealization of the experiments that have confirmed the magnetic AB effect  In the figure, an electron beam is split coherently into two spatially separated partial beams that are recombined at their ends.  Between the two partial beams is a magnetic field confined to  a region from which the electrons are excluded.  Upon being recombined, the two beams interfere with some relative phase $\Phi$.  In the reported experiments that relative phase was measured by a shift in an interference pattern.

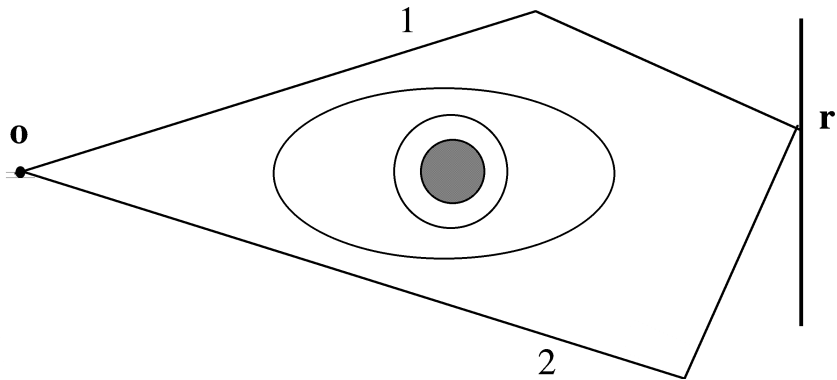

**Figure 1.**  A two-dimensional schematic view of the magnetic AB effect. The external magnetic flux is confined to a long cylinder represented by the shaded circle. Electrons are excluded from the interior of the ellipse. Point **o** is the source of the electrons.  Two typical Feynman paths, one in beam 1 and one in beam 2, are shown meeting at a point **r** of the screen where the interference  pattern is observed.

In the actual experiments the external magnetic field was provided either by a ferromagnetic whisker [6] or by a current-carrying solenoid [7,8].  In principle the field could also be provided by a rotating charged cylinder.  In any case, the external magnetic field must vanish everywhere in the figure except the interior of the shaded circle.  There must be a return flux somewhere, but that is in some region outside of figure 1 and away from the electron beams.

It follows from the Schrödinger equation that the partial wave functions in the two beams depend upon the flux $F$ in the solenoid according to

Against a proposed alternative explanation of the Aharonov-Bohm effect

$$\psi_j(F,\mathbf{r}) = \psi_j(0,\mathbf{r})\exp\left[i\frac{e}{\hbar c}\left(\int_o^\mathbf{r} \mathbf{A}(\mathbf{r}')\cdot d\mathbf{r}'\right)_j\right]. \tag{1}$$

Here the origin **o** is the point from which the two beams diverge, as shown in figure 1. **A** is the vector potential due to the flux $F$, and the subscript $j$ following the integral indicates that the path of integration is any curve confined to beam 1 or to beam 2, according to the value of $j$. When the two beams recombine to interfere at some points **r**, their relative phase is the same for all such **r** and is given by

$$\Phi(F) = \Phi(0) + \frac{e}{\hbar c}\oint \mathbf{A}(\mathbf{r}')\cdot d\mathbf{r}' = \Phi(0) + \frac{e}{\hbar c}F \tag{2}$$

In (2), the path of integration can be any closed curve that encloses the flux while remaining in the magnetic-field-free region.

Changing the flux $F$ induces a change in $\Phi$, which in turn causes a shift in the measured interference pattern. The flux dependence in (2) has been confirmed in many experiments and it is now widely accepted as showing that the action of the magnetic field on the electron is nonlocal, and more specifically that electron interacts with the local vector potential.

### 3. The proposed alternative mechanism and why it fails
The magnetic field energy inside the cylinder represented by the shaded circle in figure 1 is given by

$$W(t) = \frac{1}{8\pi}\int\left[\mathbf{B}_0(\mathbf{r})^2 + 2\mathbf{B}_0(\mathbf{r})\cdot\mathbf{B}_e(\mathbf{r},t) + \mathbf{B}_e(\mathbf{r},t)^2\right]d^3\mathbf{r}, \tag{3}$$

where $\mathbf{B}_0$ is the externally fixed field and $\mathbf{B}_e$ is the magnetic field whose source is the moving electron. The domain of **r** in the integral is the interior of the cylinder. A central idea behind the proposed alternative mechanism is that the first term in the integrand, being externally fixed, is stationary as the electron goes by but the second term is not stationary and it is proportional to the external field, so $W(t)$ is both time dependent and external-field dependent and may be relevant to the AB effect. The third term, being independent of $\mathbf{B}_0$, is irrelevant.

That reasoning is not correct. A single example suffices to reveal the source of the error. Consider the case of a long solenoid where the magnetic field is produced by the current in a resistanceless wire. Externally fixed $\mathbf{B}_0$ during the passage of the electron

Against a proposed alternative explanation of the Aharonov-Bohm effect corresponds to the limit of infinite length $\ell$ and inductance $L$ of the solenoid. For any finite $\ell$, the e.m.f. induced by the changing magnetic field whose source is the electron induces a time-dependent change in the current in the wire proportional to $1/\ell$ and therefore a change in $\mathbf{B}_0$ proportional to $1/\ell$. The first term in the energy (3) is proportional to $\ell\mathbf{B}_0^2$ and therefore has a time-dependent finite part proportional to $\mathbf{B}_0$ in the limit $\ell \to \infty$, so that first term cannot be treated as constant during the passage of the electron. The correct calculation has in fact been carried out for two specific models, one a charged rotor [2] and the other a solenoid [1], and it turned out in both cases that the sum of the first two terms in the integral is exactly constant except for terms of order $(a/\ell)$, where $a$ is the radius of the source and $\ell$ its length. The variation in time of the first term precisely cancels that of the second term. That is of course consistent with the absence of any force on the electron in the field-free region. At least in those two models, any variation of the magnetic field energy during the passage of the electron is independent of the external field.

**4. General proof**
Here I will show that any electromagnetic energy which enters the flux-bearing region must be independent of the external magnetic field, independently of all details of the source of the external field. This proof, unlike previous ones, applies directly to the experiment that used a ferromagnetic whisker as well as to the rotor or the solenoid, and it has the advantage that it eliminates irrelevant details in all cases.

Let S be the interior of the cylindrical region bounded by the larger circle in figure 1. The radius of that circle is intended to be infinitesimally greater than the radius of the flux-bearing region. As the electron passes, energy is delivered into S only by the electromagnetic field and at the rate

$$R = \frac{c}{4\pi} \int \mathbf{E}_e(\mathbf{r},t) \times \left[\mathbf{B}_0(\mathbf{r},t) + \mathbf{B}_e(\mathbf{r},t)\right] \cdot d\sigma, \tag{4}$$

where $\mathbf{E}_e$ is the electric field whose source is the passing electron, the domain of $\mathbf{r}$ in the integral is the surface of S, and $d\sigma$ is the inward-pointing normal surface area element. $\mathbf{E}_e$ and $\mathbf{B}_e$ may include waves reflected from the flux-bearing structure or other objects. Since $\mathbf{B}_0$ vanishes everywhere on the surface of S, it follows that

$$R = \frac{c}{4\pi} \int \mathbf{E}_e(\mathbf{r},t) \times \mathbf{B}_e(\mathbf{r},t) \cdot d\sigma. \tag{5}$$

That *R* may or may not vanish, but it does not depend upon the external flux. Therefore the energy variation in the magnetic field has no relevance to the Aharonov-Bohm effect.

Finally, it is necessary to remove the restriction to cylindrical geometry because in some of the experiments the magnetic field was confined to a toroidal solenoid that threaded the space between the two electron beams. In some of those experiments the external magnetic field was provided by the current in a superconducting sheath instead of in a wire. None of that creates any exception to the proof in Section 4. All that is needed is to replace the cylindrical region S by a toroidal region that encloses the toroidal solenoid.

**Acknowledgment**
This work was supported by the U.S. Department of Energy, Office of Nucl. Phys., under Contract DE-AC02-06CH11357.